\documentclass{aa}  

\usepackage{graphicx}

\usepackage{xcolor}
\usepackage{afterpage}
\usepackage{txfonts}
\usepackage[font=small]{caption}
\usepackage{geometry}
\usepackage{array}
\usepackage{makecell}
\usepackage{stfloats}
\usepackage{float}
\usepackage{multirow}
\usepackage{subcaption}
\usepackage{hyperref}
\hypersetup{
    colorlinks=true,
    linkcolor=red,
    urlcolor=cyan,
    citecolor=blue
    }
\usepackage[capitalize]{cleveref}
\usepackage{tablefootnote}
\usepackage{longtable}
\usepackage{siunitx}
\usepackage{booktabs}
\newcolumntype{P}[1]{>{\centering\arraybackslash}p{#1}}

\begin{document} 

    \title{Discovery of a probable very fast extragalactic nova in a symbiotic binary\thanks{A reproduction package for this paper is available at \tiny\url{https://doi.org/10.5281/zenodo.6817684}.}}

   \author{David Modiano and Rudy Wijnands}
   \institute{Anton Pannekoek Institute for Astronomy, University of Amsterdam, Postbus 94249, 1090 GE Amsterdam, The Netherlands\\
   \email{d.modiano@uva.nl}
   }
 
  \abstract
 {Very fast novae are novae which evolve exceptionally quickly (on timescales of only days). Due to their rapid evolution, very fast novae are challenging to detect and study, especially at early times. Here we report the discovery, which was made as part of our Transient UV Objects project, of a probable very fast nova in the nearby spiral galaxy NGC 300. We detected the rise to the peak (which is rarely observed for very fast novae) in the near-ultraviolet (NUV), with the first detection just $\sim$ 2 hours after the eruption started. The peak and early stages of the decay were also observed in UV and optical bands. The source rapidly decayed (2 NUV magnitudes within 3.5 days), making it one of the fastest novae known. In addition, a likely quiescent counterpart was found in archival near-infrared Spitzer and VIRCAM images but not in any deep optical and UV observations, indicating a very red spectral shape in quiescence. The outburst and quiescence properties suggest that the system is likely a symbiotic binary. We discuss this new transient in the context of very fast novae in general and specifically as a promising supernova Type Ia progenitor candidate, due to its very high inferred WD mass ($\sim$ 1.35 M$_\odot$; determined by comparing this source to other very fast novae).}

   \keywords{methods: data analysis -- methods: observational -- ultraviolet: stars -- binaries: symbiotic -- novae}
   
    \titlerunning{Discovery of a very fast nova}
    \authorrunning{Modiano \& Wijnands.}
   
   \maketitle

\section{Introduction}  \label{introduction}
\interfootnotelinepenalty=10000

White dwarfs (WDs) in binary systems can accrete matter from their companions. If the companion is a main-sequence (MS) star (although in rare cases it can be another WD), then mass transfer occurs via Roche-Lobe overflow and the system is usually known as a cataclysmic variable (CV; for an extended review see \citealp{Warner_1995}). If the donor is an evolved giant, the orbital separation is large (at least hundreds of days) and the system is known as a symbiotic binary (or a symbiotic; see \citealp{Mikolajewska_2012,Munari_2019} for reviews). In this case, the WD accretes mass from the strong stellar wind of the companion, although for symbiotics with short orbital periods ($\lesssim$600 days) Roche-Lobe overflow may also occur (and could be the primary method of accretion; \citealp{Mikolajewska_2010}) .\

When Roche-Lobe overflow is the dominant mode of accretion (i.e. in CVs and short-period symbiotics), then the system can undergo explosive thermonuclear outbursts known as novae \citep{Gallagher_1978,Chomiuk_2021}. These eruptions arise because a hydrogen-rich layer of accreted material accumulates on the WD surface, causing the temperature and density to increase and eventually reach the critical conditions for unstable nuclear burning \citep{Starrfield_1972,Bode_2008}. This triggers a thermonuclear runaway event (TNR), which causes the accreted envelope to rapidly expand, and consequently the system to brighten in the ultraviolet (UV) and optical by up to a factor of 1 million (8--15 visual magnitudes) on timescales of days to weeks (for reviews see \citealp{Bode_2008,Starrfield_2016,Chomiuk_2021}). \

Novae are often classified according to their light curve properties, in particular the timescale of their decay. The `speed class' is commonly quantified by the parameters $t_{2}$ and $t_{3}$, which indicate the time taken for the brightness of the nova to decline from its peak by 2 and 3 visual magnitudes, respectively. The classes range from `very fast' ($t_{2}$ < 10 d) to `very slow' ($t_{2}$ > 150 d; \citealp{payne1964galactic}). Despite some caveats (see Sect. \ref{discussion}), this parameter is important because the speed of decay is linked to the WD mass, with the fastest-evolving novae likely to harbour the most massive WDs ($\sim$1.4 M$_{\odot}$; \citealp{Hachisu_2016,Shara_2017,Hachisu_2018,Darnley_2021}). Thus, very fast novae are of interest because of their extreme nature both in terms of light curve properties (and the underlying physical processes) and WD mass. \

So far, a few tens of very fast novae have been discovered, but with only seven originating in symbiotics\footnote{\tiny Here is it worth making a brief note on terminology with regards to novae. The term `symbiotic novae' is usually reserved for very slow (years to decades) and low-amplitude (2--6 magnitudes) outbursts that occur in typical, wide-orbit (orbital periods $\gtrsim$800 days) symbiotics \citep{Allen_1980}. Due to the physical properties of these slow novae (low accretion rates, low WD masses, and wind-only accretion) and their observational signatures (very slow and very faint), it is important to distinguish them from all other novae. Thus, novae originating in symbiotics are referred to as such, and not as `symbiotic novae'. In this paper, we do not include the latter when discussing novae.} (\citealp{Schaefer_2010,Darnley_2016,Jencson_2021} and references therein). There are still many uncertainties regarding the mechanisms of these remarkable outbursts, such as the temperature evolution (e.g. \citealp{Darnley_2016}) and the geometrical structure of the expanding envelope (e.g, \citealp{Cao_2012}). In addition, many very fast novae exhibit plateaus in their light curve (e.g. \citealp{Strope_2010}) which are also not well understood (possibly they arise from the irradiation of the accretion disc by emission from the WD; \citealt{Hachisu_2000_2}). \

Very fast novae are additionally noteworthy because they have been proposed as progenitors of Type Ia supernovae (SNe; \citealp{Maoz_2014,Hillman_2016,Hachisu_2018}). Although there is general agreement in that Type Ia SNe result from the thermonuclear 
detonation of a WD after its mass has increased to the Chandrasekhar limit ($\sim$1.4 M$_{\odot}$; \citealp{Nomoto_1982}), it is not yet clear how this is achieved. Very fast novae may provide an efficient pathway to sufficiently increasing the mass of a WD, due to their high WD masses and the fact that they can grow \citep{Hillman_2016}. A particularly likely channel may be the very fast novae which originate in symbiotics (e.g. \citealp{Patat_2011}). However, although not precisely known, the rates of very fast novae appear to be too low to account for all Type Ia SNe (although this is true of all the likely progenitor channels, suggesting that there are multiple channels; see e.g. \citealp{Schaefer_2010}). \

Recently, we have initiated the Transient UV Objects (TUVO) Project, through which we search for and study transients in the relatively underutilised UV wavelength range. Within this project, we focus on the UV data obtained by the Ultraviolet and Optical Telescope (UVOT, \citealp{Roming_2005,Breeveld_2010}) aboard the \textit{Neil Gehrels Swift Observatory} (\textit{Swift}, \citealp{Gehrels_2004}). Through our specialised pipeline used to analyse these UVOT data \citep{Modiano_2022}, we regularly detect serendipitous new and archival transients. Since novae exhibit strong UV outbursts, this method is well-suited to the discovery and study of these systems. In this letter, we present the TUVO discovery and a photometric study of TUVO-22albb, a new UV transient we detected in the nearby spiral galaxy NGC 300 (1.88 Mpc; \citealp{Gieren_2005}). We determine it to be most likely one of the rare, very fast novae originating in symbiotics, and find that it is one of the fastest-evolving novae known.

\section{Observations} \label{observations}

\subsection{Discovery} \label{discovery}

\begin{figure}
    \centering
    \includegraphics[width=0.5\textwidth]{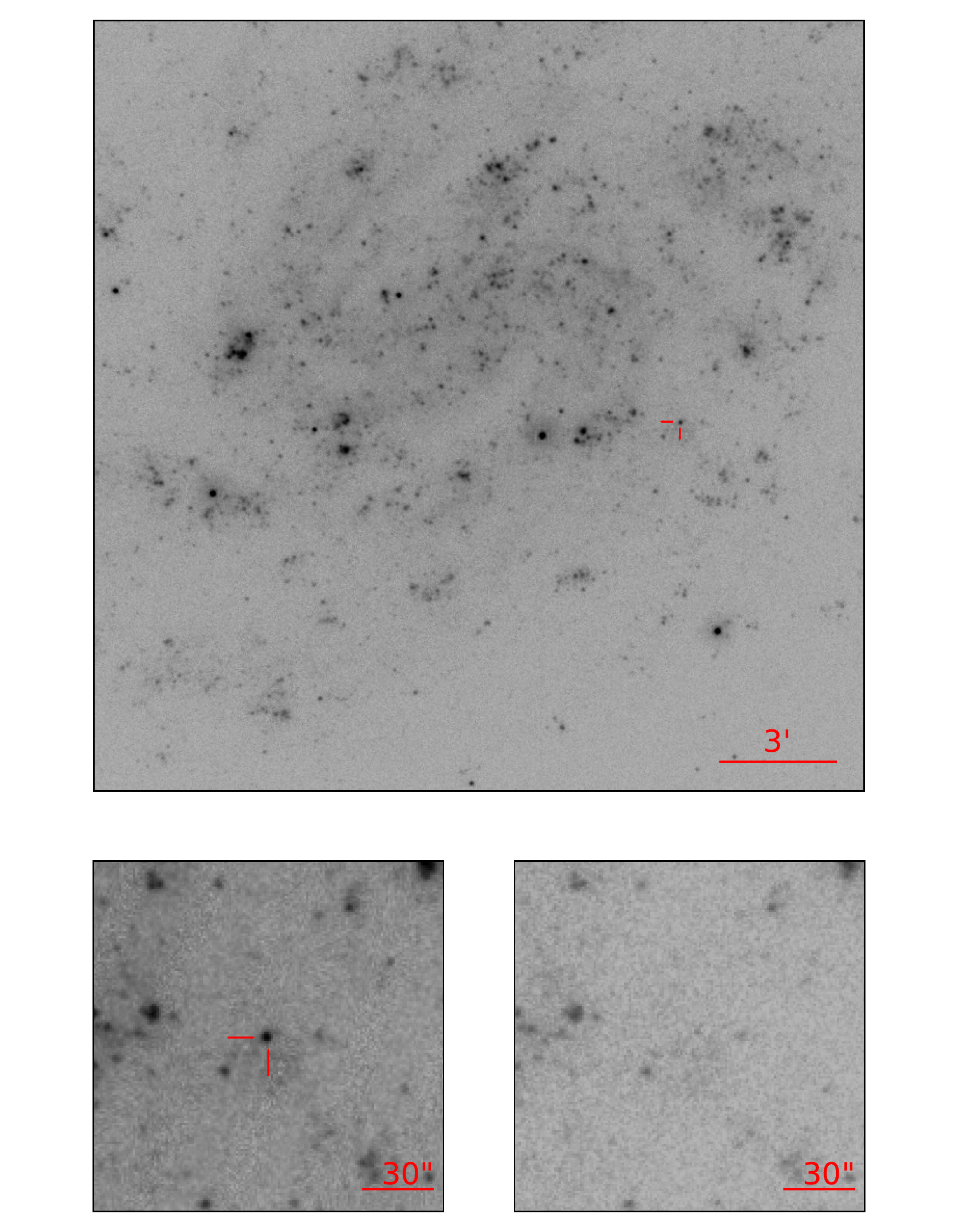}
    \caption{Top panel: UVOT $UVW1$ exposure of NGC 300 in which TUVO-22albb was discovered (Observation ID 00030848003, exposure 15). Bottom left panel: zoomed-in view of the top panel image, but centred on TUVO-22albb. The transient is labelled. Bottom right panel: $UVW1$ reference image during quiescence (Observation ID 00031210007, exposure 1). North is up and east is left.}
    \label{fig:uvot_images}
\end{figure}

Using our pipeline, we detected a transient in an archival\footnote{\tiny Archival \textit{Swift} data are hosted at the High Energy Astrophysics Science Archive Research Centre (HEASARC); at \url{https://heasarc.gsfc.nasa.gov/cgi-bin/W3Browse/swift.pl}.} UVOT exposure of NGC 300, taken using the $UVW1$ filter (central wavelength $2600$ $\AA$; \citealp{Poole_2008}). The source was denoted as TUVO-22albb. We report the position at right ascension=00:54:34.26 and declination=-37:42:50.7, with an error circle of \SI{0.5}{\arcsecond} (due to the UVOT astrometric solution; see \citealp{Poole_2008}). In Fig. \ref{fig:uvot_images} we show the discovery image, which was taken on Dec. 31, 2006, and cutouts of the discovery and reference images zoomed in on the transient, clearly showing the transient only visible in the discovery image.\ 

A preliminary light curve was automatically created by our pipeline using all archival UVOT data covering the position of the source (shown in Appendix \ref{appendix_pv}). This data set includes 164 observations, and as per the UVOT data structure\footnote{\tiny\url{https://www.swift.ac.uk/analysis/uvot/files.php}}, each observation consists of one or multiple individual exposures per filter. The full data set spanned $\sim$16 years, was of highly varying cadence (between hours and years), and contained data in all six UVOT filters\footnote{\tiny\url{https://www.swift.ac.uk/analysis/uvot/filters.php}}. This long-term light curve indicated that the source was only detected during a $\sim$ 10 day period, so we selected these exposures with which to carry out a more customised analysis (see Appendix \ref{appendix_coadd} for details of the observations used). We note that the late stages of the outburst may have continued for several weeks, but there were no UVOT observations in this period. The gap between the last detection and the subsequent non-detection is $\sim$20 days.

\subsection{Outburst photometry} \label{outburst_photometry}

\begin{figure}[t!]
    \centering
    \includegraphics[width=0.48\textwidth]{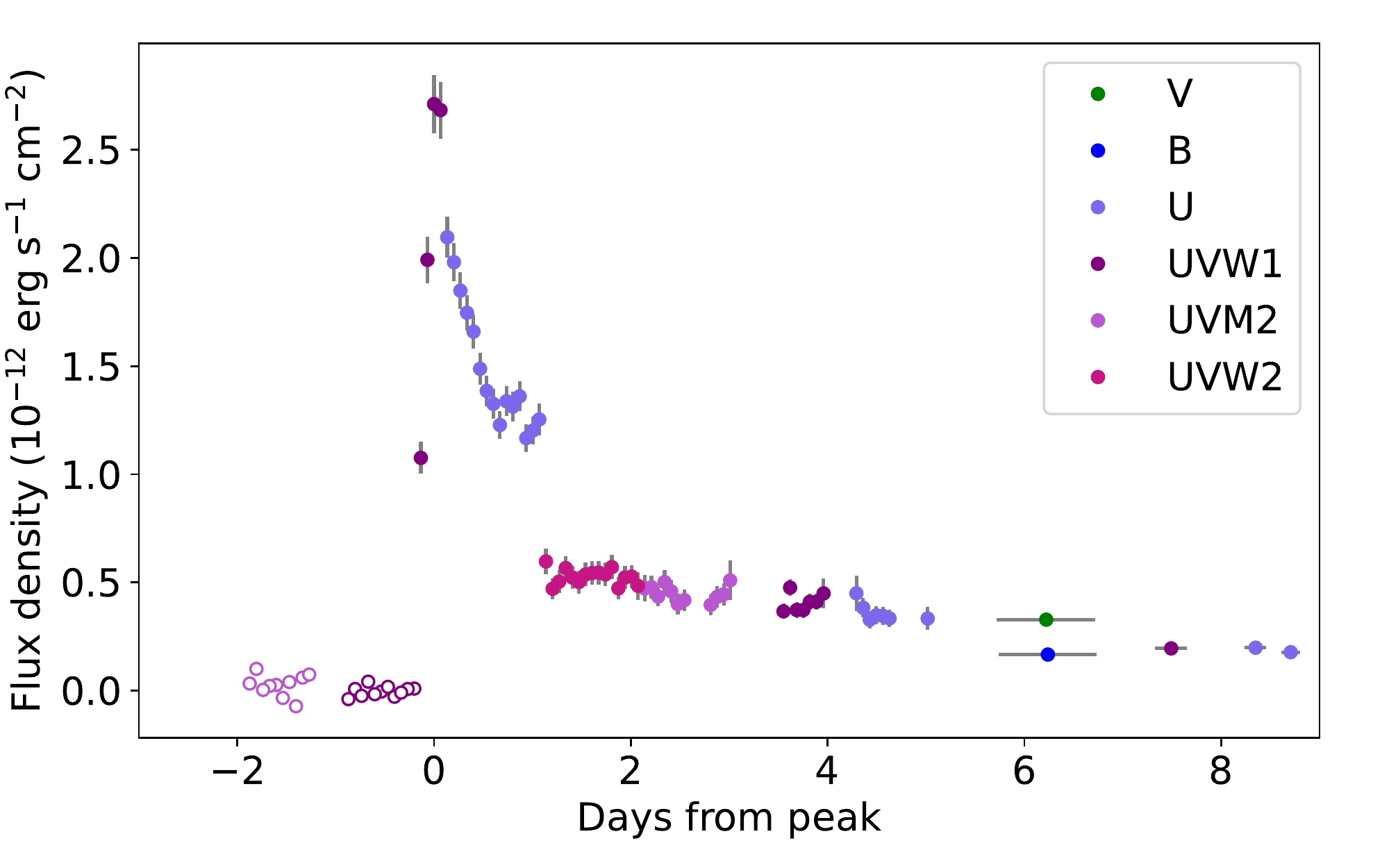}
    \caption{Outburst light curve of TUVO-22albb. Day zero is defined as MJD=54099.9, which is the time of the exposure during the peak of the outburst. Data points with error bars on the time indicate exposures that were co-added, and open circles indicate exposures in which the source was not detected (see Appendices \ref{appendix_coadd} and \ref{appendix_photometry} for details).}
    \label{fig:lightcurve}
\end{figure}

To characterise the outburst brightness evolution of the transient, we used all the selected outburst exposures to construct a light curve, which we display in Fig. \ref{fig:lightcurve}. Before performing the photometry, we inspected all the exposures during the outburst to check for secure detections of the source. We noted that during the later stages of the outburst ($\gtrsim$ 6 days after peak), the source had dimmed such that it was not clearly detectable in every individual exposure. Therefore, for these data we co-added some exposures (see Appendix \ref{appendix_coadd} for details). \ 

We then carried out photometry on all the individual or resulting co-added exposures. All fluxes were obtained using the standard UVOT photometry tool \texttt{uvotsource}\footnote{\tiny\url{https://heasarc.gsfc.nasa.gov/lheasoft/ftools/headas/uvotsource.html}}. The methods used for photometry, including aperture selection and dealing with diffuse background emission, are described in Appendix \ref{appendix_photometry}.  The fluxes were corrected for extinction using the Galactic dust and extinction maps available at NASA/IPAC Infrared Science Archive\footnote{\tiny\url{https://irsa.ipac.caltech.edu/applications/DUST/}}. Extinction in the line of sight of TUVO-22albb is small ($E_{B-V}=0.01$), so this had only a small effect (typically $\sim$5 \%) on the light curve. We note that the $U$, $UVW1$, and $UVW2$ UVOT filter transmission is contaminated by a red leak. However, several studies (e.g. \citealp{Brown_2010,Siegel_2012,Siegel_2014}) have found that this does not have a significant effect for hot ($\gtrsim$5000--10\ 000 K) sources such as novae at early times (see Sect. \ref{discussion}), so we did not correct for it. For easier comparison with other similar sources for which light curves are often provided in magnitudes, the final corrected fluxes were also converted back into (AB) magnitudes; these are provided in Appendix \ref{appendix_photometry} (Table \ref{tab:appendix_mags_final}) for all observations in which a detection was made.

\section{Analysis and results}  \label{analysis_results}

\subsection{Outburst: light curve properties} \label{light_curve_properties}

The rise and peak of the outburst were observed only using the $UVW1$ filter; the initial rapid decline was observed only using the $U$ filter; and the slower subsequent decay was observed using all six UVOT filters. Unfortunately, observations between the filters do not overlap in time (except the late $B$ and $V$ observations), so we cannot precisely discern the profile of the entire light curve. However, at maximum brightness the spectral energy distribution (SED) of this source is likely strongest in the $U$ or near-UV (NUV; e.g. \citealp{Jencson_2021}), so a reasonable assumption is that the SED between the $U$ and $UVW1$ bands (which are only separated by $\sim$ 800 $\AA$) at the peak was fairly flat. Since these were the two filters used for the exposures during and immediately after the light curve peak, the early light curve profile is reasonably well approximated by the combination of the early $UVW1$ and $U$ data. Overall, the outburst light curve (Fig. \ref{fig:lightcurve}) of the transient can be broadly characterised by a very rapid (hours) rise, followed by a initial steep decay until t~$\sim$~0.6 d and then a subsequent plateau and slower decay phase (where we have defined t = 0 as MJD=54099.9, which is the time of the peak of the light curve in the $UVW1$ filter).

\begin{figure*}[t]
    \centering
    \includegraphics[width=1.0\textwidth]{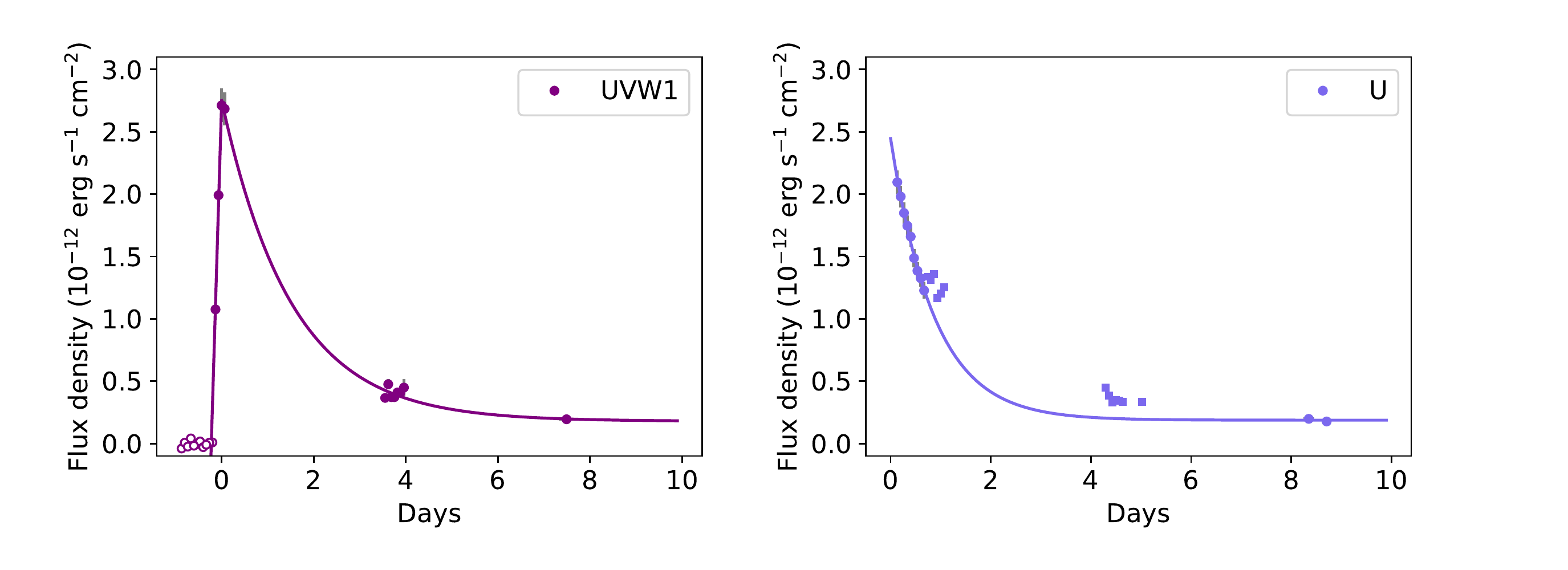}
    \caption{Fits to the $UVW1$ (left panel) and $U$ (right panel) light curves. Detections are shown as solid circles and non-detections are shown as open circles. The rise in $UVW1$ is fit with a straight line using the first three detections. The decays for both $U$ and $UVW1$ are fit with exponential functions. For $U$, the data shown with squares were not included in the fit (see Sect. \ref{light_curve_properties} for details). For $UVW1$, all detections (starting from the peak) were used in the fit. The inferred $t_{2}$ from the fit for $UVW1$ is 3.5 days, just before the cluster of data.}
    \label{fig:lc_exp_fits}
\end{figure*}

After $\sim$1 day, a drop in flux by a factor of 2 occurs between the $U$ and $UVW2$ band data but this happens in less than 2 hours. This suggests that it does not represent a real decrease in overall luminosity but that the source was fainter in $UVW2$ than in $U$, indicating an SED peaking towards the NUV or $U$. Therefore, the assumption of a flat SED between the $UVW1$ and $U$ filters during the first day of the outburst is not valid for the entire SED (and indeed is only a first-order approximation). However, for the remainder of the light curve (i.e. during the slow decline phase starting from day $\sim$2) the very similar fluxes and proximity (in time) of observations between the different filters suggest that the SED rapidly switched to exhibiting a fairly flat shape across the full UV-optical range. \

The start time of the eruption can be very well constrained by the non-detections in the data obtained using the $UVW1$ filter just a few hours before the first detection (using the same filter). We approximated the rise with a straight-line fit to the earliest three $UVW1$ detections (Fig. \ref{fig:lc_exp_fits}, left panel). This gave a flux of zero at 5.3 hours (0.22 days) before the peak, which gives a good indication of the rise time. 

The decay stage is more complex than a simple exponential decay. Nonetheless, exponential fits can still be illustrative of the source behaviour, and allow us to estimate $t_{2}$. Therefore, we fitted an exponential decay function to the $U$ rapid-decline data (up to $\sim$0.6 days) together with the late-time detections at >8 days. The fit is shown in Fig. \ref{fig:lc_exp_fits} (right panel). We found the best-fit $\tau_{U}$ = -1.15 $\pm$ 0.04 d$^{-1}$. Due to the decline break, we cannot fit the decay in $U$ before and after 0.6 days with the same exponential function, so we cannot constrain $t_{2,U}$ well. However, in order to quantitatively compare the fast initial and slower later declines, we also fitted an exponential function to the data omitting the steep initial decline, obtaining a much slower $\tau_{U}$ = -0.48 $\pm$ 0.04 d$^{-1}$. Assuming the $U$ peak brightness was similar to that for $UVW1$, an upper limit for $t_{2,U}$ can be estimated as 4.3 days, which is the time between the peak and the first $U$ exposure for which the brightness had declined by more than 2 magnitudes. \ 

We also fitted an exponential decay function to the $UVW1$ data, displayed in Fig. \ref{fig:lc_exp_fits} (left panel), with the best fit giving $\tau_{UVW1}$ = -0.66 $\pm$ 0.04 d$^{-1}$. This indicates slower decay than that inferred from the t $\leq$ 0.6 day $U$ data, which reflects a combination of the faster early decay before 0.6 days and shallower subsequent decay. The $t_{2,UVW1}$ value we infer from the exponential fit is 3.5 days\footnote{\tiny Since the peak is clearly observed in $UVW1$, a robust upper limit for $t_{2}$ independent of the fitting can be determined by calculating the time between the peak and the $UVW1$ exposure during which the brightness had decreased by 2.2 magnitudes. This gives $t_{2}$ < 3.55 days, which is very close the the value of 3.5 days inferred from the exponential fit.}. There is not sufficient sampling in the $UVW1$ band for it to be useful to fit the light curve with two different exponential functions. Due to the brightness of the background emission at the source location, the photometry never exhibits a 3 magnitude decrease from the peak, so we are unable to determine $t_{3}$. However, at t = 8.3 days it has declined by $\sim$ 2.9 magnitudes, so we take this as a lower limit for $t_{3,UVW1}$. \

\subsection{Quiescence} \label{results_quiescence}

The long-term UVOT light curve (Fig. \ref{fig:pv_appendix}) did not reveal any detections outside of the outburst period, indicating that only one outburst was observed by UVOT. To constrain the source brightness during quiescence, we co-added all pre- and post-outburst UVOT exposures per filter, which ranged in time from 2007 to 2022 (see Appendix \ref{appendix_pv}). This resulted in data sets with total exposure times of $\sim$17 ks, $\sim$23 ks, $\sim$66 ks, $\sim$91 ks, $\sim$91 ks, and $\sim$98 ks in the filters $V$, $B$, $U$, $UVW1$, $UVM2$, and $UVW2$, respectively. The source was not detected in any of the co-added exposures. While the detection limits of these exposures overall reached 23-24 magnitudes, the diffuse emission due to NGC 300 at the source position resulted in weaker upper limits for the quiescent counterpart. To determine these, we ran \texttt{uvotsource} at the source position (when no source is detected the tool calculates a 3-$\sigma$ upper limit). We obtained upper limits of 18.6, 18.7, 19.5, 20.5, 20.7, and 20.5 magnitudes in the filters $V$, $B$, $U$, $UVW1$, $UVM2$, and $UVW2$, respectively (see Appendix \ref{appendix_photometry}). Comparing the $U$ and $UVW1$ quiescence upper limits to their corresponding peak magnitudes of $\sim$17.9, this suggests outburst amplitudes of at least 1.6 and 2.6 magnitudes, respectively. However, during quiescence the source was likely much fainter than the upper limits (see Sect. \ref{discussion}) so these are very weak lower limits to the amplitudes.\ 

We searched various publicly available astronomical databases for archival observations covering the position of the source (see Table A.1 in \citealp{Modiano_2022} for details of all the catalogues we probe automatically in the TUVO project). We found no previous classifications of this source in any of these databases, and no detections in X-ray, UV, or optical catalogues. We looked at archival optical ($g$ band; 380--550nm) images from OmegaCam \citep{Kuijken_2011} mounted on the ESO-VLT Survey Telescope \citep{Shanks_2015} and did not detect the source, with an upper limit of $\sim$ 23.4 magnitudes (5-$\sigma$; the provided detection limit). This significantly improves the optical upper limits on the quiescent counterpart determined from UVOT.\

Potential matches to our source, which we define as any catalogued source within \SI{1}{\arcsecond} of the position of TUVO-22albb, were found in archival near-infrared (NIR) images obtained by the Spitzer \citep{Werner_2004} Infrared Array Camera (IRAC; \citealp{Fazio_2004}) and by VIRCAM\footnote{\tiny\url{https://www.eso.org/sci/facilities/paranal/instruments/vircam/inst.html}} mounted on the VISTA telescope (\citealp{Sutherland_2015}; source IDs SSTSL2 J005434.20-374251.3 and 473505681893, respectively). The Spitzer-IRAC catalogued photometry measurements for this source in Channel 1 (3.1--3.9 $\mu$m) and Channel 2 (3.9--5.1 $\mu$m) were 17.9$\pm$0.1 and 18.2$\pm$0.1 magnitudes, respectively. The VISTA-VIRCAM $J$ (1.15--1.4 $\mu$m) and $K_{s}$ (1.95--2.35 $\mu$m) band photometry measurements were 19.1$\pm$0.1 and 18.2$\pm$0.2 magnitudes, respectively. \ 

\subsection{X-ray data} \label{xrt_analysis}

We also examined the data taken with the X-ray telescope (XRT; \citealp{Burrows_2005}) aboard \textit{Swift}, which operates simultaneously with the UVOT. Combining both quiescence and outburst observations did not result in a detection at the source position (see Appendix \ref{appendix_xrt} for details). However, the observation subsequent to the last detection in the light curve, which was taken $\sim$30 days after the peak, possibly revealed a very faint X-ray source (albeit with SNR $\sim$ 1.9) coincident with the position of TUVO-22albb. Using the \texttt{ximage}\footnote{\tiny\url{https://www.swift.ac.uk/analysis/xrt/xrtcentroid.php}} tool, we found a count rate of 0.004$\pm$0.002 c s$^{-1}$, and we used the Webpimms\footnote{\tiny\url{https://heasarc.gsfc.nasa.gov/cgi-bin/Tools/w3pimms/w3pimms.pl}} tool available from HEASARC to calculate a flux in the 0.3-10 keV band of 1.9$\times$10$^{-13}$ erg cm$^{-2}$ s$^{-1}$ (for details see Appendix \ref{appendix_xrt}). Assuming the distance to NGC 300 of 1.88 Mpc, this corresponds to an X-ray luminosity of 8.1$\times$10$^{37}$ erg s$^{-1}$.

\section{Discussion} \label{discussion}

Using our method to search for UV transients in \textit{Swift} UVOT data \citep{Modiano_2022}, we discovered an archival transient in the nearby spiral galaxy NGC 300. The UV and optical light curve was exceptionally well sampled, including detections during the rise, peak, and early (< 10 days) decay stages. We used the light curve to characterise the evolution of the outburst. We further investigated the quiescent counterpart using both \textit{Swift} data and additional publicly available optical and NIR images. \

\begin{figure}[h]
    \centering
    \includegraphics[width=0.48\textwidth]{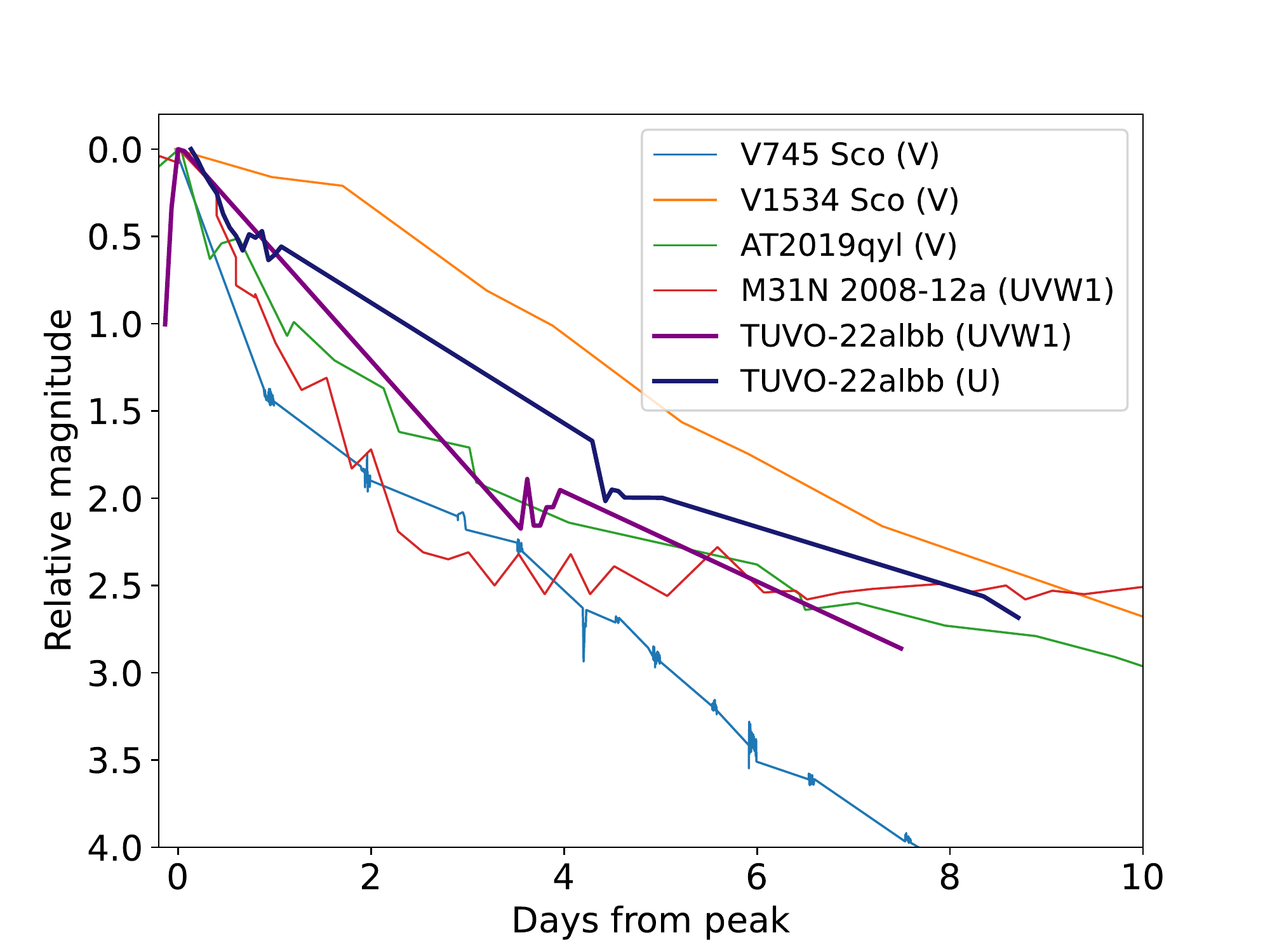}
    \caption{Light curves of TUVO-22albb and four well-known very fast novae (for which data are available), all of which likely originate in symbiotics. The TUVO-22albb light curves are shown with thicker lines. All light curves have been shifted such that the time of the peaks is at day zero and the magnitude of the peaks is set to zero. The data for M31N 2008-12a were obtained from the UVOT data archive and processed using our pipeline (see Sect. \ref{discovery} and \citealp{Modiano_2022}); for the other sources their data were obtained from the American Association of Variable Star Observers (\url{https://www.aavso.org/}).}
    \label{fig:lc_comparison}
\end{figure}

The data presented suggest that TUVO-22albb is most likely a very fast nova originating in a symbiotic. The arguments in favour of this classification are:
\begin{itemize}
     \item The spatial coincidence with NGC 300. This suggests an association with this galaxy and that it is not a fainter, foreground Milky Way transient such as a stellar flare or a dwarf nova (DN; a 2--6 mag outburst caused by accretion disc instabilities in a CV; see \citealp{Lasota_2001}). We note that flares are further ruled out by the transient duration: flares last between a seconds and $\sim$ 1 day (e.g. \citealp{Shibayama_2013,Gunther_2020}), while this transient was active for at least 9 days. If the association with NGC 300 is correct, this implies a distance of 1.88 Mpc \citep{Gieren_2005} and consequently peak absolute (AB) magnitudes of $M_{UVW1}$ = $-$8.3 and $M_{U}$ $\leq$ $-$8.4. This is consistent with typical nova luminosities (see e.g. \citealp{Schaefer_2010,Strope_2010}).
    \item The shape and speed of evolution of the light curve. The rapid rise, occurring in just $\sim$5 hours, and the fast decay in the subsequent days, closely resemble that of other very fast novae (e.g. \citealp{Darnley_2016,Jencson_2021}). A plot displaying the light curves of some very fast novae originating in symbiotics together with that of TUVO-22albb is displayed in Fig. \ref{fig:lc_comparison} (see below for a more extended discussion comparing TUVO-22albb to similar sources). The $t_{2}$ of $\sim$3.5 days places TUVO-22albb in the speed class of very fast novae and it is one of the fastest known novae.
    \item The evolution of the SED. The data are consistent with the SED peaking in the U or NUV at early times (t < 1 day) and then flattening. For very fast novae, it is expected that the material ejected expands and cools, therefore shifting the SED from a hot, blue continuum to a flatter SED (e.g. \citealp{Darnley_2021}; see also the extended discussion below).
    \item The brightness and colour of the likely quiescent counterpart we found in NIR images. For this source $J$-$K_{S}$ = 0.9 and $M_{Ks}$ = $-$8.2 (at the distance of NGC 300). Using the colour-magnitude diagrams of evolved giants from \citet{Cioni_2006} and \citet{Whitelock_2009}, we find that this source is brighter than the tip of the red giant (RG) branch ($M_{Ks}\sim-6$), and consistent with asymptotic giant branch (AGB) stars. The likely presence of an evolved giant indicates that the system is a symbiotic, and specifically we suggest that the companion in TUVO-22albb is an AGB star.
    \item The lack of detections in archival optical and UV images down to $\sim$23.4 magnitudes. This further rules out the source being a nearby (Milky Way) DN, because its quiescent optical brightness should have been < 23 mags (moreover, the quiescent SED in DNe is expected to be flat, not very red). Similarly, stellar flares do not cause brightenings of more than a few magnitudes (e.g. \citealp{Gunther_2020}). Unfortunately, the Hubble Space Telescope did not image the exact position of our source (this would have provided a very good constraint on the quiescence magnitude and therefore the outburst amplitude).
    \item The possible X-ray source coincident with the position of TUVO-22albb in an XRT image taken $\sim$30 days after the outburst peak. The luminosity of the source ($\sim$8$\times$10$^{37}$ erg s$^{-1}$; assuming it was a real detection) is similar to that of other X-ray detections of novae days to weeks after the peak. For example, for the very fast nova U Scorpii soft X-rays were detected and peak in brightness $\sim$30 days after the optical maximum \citep{Mason_2012}. Similar X-ray detections were made for very fast novae in symbiotics (peaking at varying times after the optical nova peak), such as V745 Sco \citep{Page_2015}, V3890 Sgr \citep{Page_2020}, and RS Oph \citep{Page_2022}. X-ray detections are likely influenced by the orbital phase during the observations (e.g. \citealp{Mason_2012}).
    
\end{itemize}

Although the $t_{2}$ parameter does not encapsulate the many complications and diversities exhibited by nova light curves, it is still a good indication of the speed class and can be used to infer properties of the system. The $t_{2,NUV}$ of 3.5 days makes TUVO-22albb one the fastest-evolving novae known. This source then belongs to the class of very fast novae originating in symbiotics. To our knowledge, until now only eight of these sources were known: the famous RS Ophiuci (e.g. \citealp{Munari_2022}), along with T CrB \citep{Ilkiewicz_2016}, V745 Sco \citep{Sekiguchi_1990}, V1534 Sco \citep{Joshi_2015}, V3890 Sgr \citep{Page_2020}, V407 Cyg \citep{Munari_2011}, M31N 2008-12a \citep{Darnley_2016}, and the recently discovered AT 2019qyl \citep{Jencson_2021}. Among these, TUVO-22albb together with AT 2019qyl exhibit the third fastest $t_{2}$ (3.5 days; although the fastest, V745 Sco, was not observed at peak, so its $t_{2}$ of 2.0 days is uncertain). Fig. \ref{fig:lc_comparison} shows the light curves of TUVO-22albb along with four other very fast novae originating in symbiotics for which data are available. \ 

The shape and speed of evolution of the light curve of TUVO-22albb clearly resembles those of other very fast novae in symbiotics. The decline is slightly slower than those of M31N 2008-12a and V745 Sco, slightly faster than that of V1534 Sco, and very similar to that of AT 2019qyl (as also suggested by the same $t_{2}$ of 3.5 days). The shape of the decay is also similar. We note that due to the inhomogenous sampling for each particular filter for TUVO-22albb, the light curve shape suggested by the connected lines is slightly misleading; most likely the true decline is exponential-like (see Sect. \ref{analysis_results}) and therefore matches the behaviour of AT 2019qyl even more clearly. We note that this comparison is based on different filters (due to data availability). However, the general behaviour of the decay in very fast novae light curves is typically very similar between filters in the optical to NUV; therefore comparisons of the decay shapes still give insightful information about similarities between different systems. \ 

In Table \ref{tab:comparison_table} we compile basic properties (the absolute magnitude at maximum brightness $M_{peak}$, $t_{2}$, $t_{3}$, and the most likely companion type) of all the known very fast novae which are thought to originate in symbiotics (to our knowledge). By $t_{2}$ and $t_{3}$, TUVO-22albb ($t_{2}$ = 3.5 d and $t_{3}$ > 8.3 d) is slightly faster than the averages of $\sim$4.4 d and $\sim$10.3 d, respectively (where the averages include TUVO-22albb). The absolute magnitude of TUVO-22albb at its peak shows that it is on the brighter side of similar sources, though our observations at the peak were in the NUV ($U$ and $UVW1$) while other sources typically have only $V$-band observations at the peak. The table further highlights the remarkable similarity of this source with AT 2019qyl, in terms of peak brightness, decay properties, and most likely companion stars. \

\begin{table}[h]
    \centering
    \small
    \caption{Comparison table of properties of all the known (to our knowledge) very fast novae thought to originate in symbiotics.}
    \begin{tabular}{ccccc}
    \toprule
        Object & $M_{peak}$ (filter) & $t_{2}$ & $t_{3}$ & Donor \\ \hline \hline
        RS Oph & $-$8.0 ($V$) & 7 & 14 & RG \\
        V745 Sco & $-$7.9 ($V$) & 2 & 4 & RG \\
        V1534 Sco & $-$5.8 ($V$) & 5.6 & 9.2 & RG \\
        T CrB & $-$6.9 ($V$) & 4 & 6 & RG \\
        V407 Cyg & $-$9.0 ($V$) & 5.9 & 24 & Mira \\
        M31N 2008-12a & $-$6.0 ($V$) & 1.65 & 2.47 & RG \\
        V3890 Sgr & $-$5.2 ($V$) & 6 & 14 & RG \\
        AT2019qyl & $-$9.2 ($V$), -9.8 ($U$) & 3.5 & 10.3 & AGB \\
        TUVO-22albb & $-$9.2 ($U$) & 3.5 & >8.3 & AGB \\
    \bottomrule
    \end{tabular}
    \parbox{0.45\textwidth}{\small%
    \vspace{1eX}
    \textbf{Notes.} The data for M31N 2008-12a are taken from \citet{Darnley_2016}; for V3890 from \citet{Strope_2010}; for AT2019qyl from \citet{Jencson_2021}; for TUVO-22albb from this paper; for all others from \citet{Hachisu_2018} and references therein. As novae magnitudes are usually provided in the Vega system, all magnitudes in this table (including for TUVO-22albb) are given in the Vega system (this is in contrast with the AB system which we use for other magnitudes quoted this paper).}
    \label{tab:comparison_table}
\end{table}

Here we highlight a limitation in comparisons using $t_{2}$. A striking feature of the light curve of TUVO-22albb is the sharp break in the decline at $\sim$0.6 days, which marks the start of a plateau phase. Such plateaus have been observed in some novae, mostly fast novae with short recurrence times. Indeed, this feature has been suggested as a telltale signature for such systems (\citealp{Schaefer_2010,Darnley_2016}) and thus strengthens our interpretation of TUVO-22albb as a member of this class of object. The tendency is that the faster the nova, the earlier the plateau phase begins (e.g. \citealp{Strope_2010}); however, to our knowledge no nova exhibits a plateau beginning as early as 0.6 days. For instance, the plateau phase for M31N 2008-12a (one of the fastest novae known, with a similar light curve to TUVO-22albb but with $t_{2}$ < 2 days) begins at $\sim$4 days. This highlights the caveats associated with using the $t_{2}$ parameter: for most very fast novae, the plateau phase (if present) occurs after $t_{2}$ has passed. Therefore, when comparing the early decay speed of TUVO-22albb to that of other very fast novae, our reported $t_{2}$ is artificially high (i.e. slow). Extrapolating the early rapid decay observed in $U$ (i.e. ignoring the plateau data; see Fig. \ref{fig:lc_exp_fits}, right panel) results in a faster $t_{2}$ of 2.2 days. \

Plateaus in nova light curves are thought to be caused by the irradiation of the disc by the radiation emitted from the nuclear burning on the WD surface (which becomes visible as the envelope recedes), combined with the decreasing flux from photons emitted in the ejecta \citep{Schaefer_2010,Hachisu_2000}. \citet{Hachisu_2000_2} modelled fast nova light curves and find that several properties of the system, including the accretion disc radius, the thickness of the disc, and the irradiation efficiency might affect the plateau. However, it is not clear what would cause the very early plateau phase of TUVO-22albb.\

The SED evolution of TUVO-22albb is difficult to discern because of the lack of simultaneous data in multiple filters. Nonetheless, as suggested in Sect. \ref{light_curve_properties}, at the peak of the outburst the SED is consistent with having its maximum in the $U$ and NUV followed by a rapid (a few days) evolution towards a flat spectrum across the UV-optical during the decay of the outburst. This is similar to the SED evolution for both AT 2019qyl and M31N 2008-12a, where the SEDs are observed to peak in the NUV at maximum brightness followed by a flattening of the SED over the course of the first $\sim$ 10 days (see Fig. 5 in \citealp{Jencson_2021} for AT 2019qyl and Fig. 16 in \citealp{Darnley_2016} for M31N 2008-12a). Flat UV-optical SEDs are expected throughout the outburst for very fast novae in symbiotics but not those in CVs \citep{Darnley_2016,Hachisu_2016}. These similarities therefore further suggest a symbiotic nature for TUVO-22albb. Rapid SED evolution is indicative of a fast evolution of the temperature, which is achievable with a relatively low mass ejecta shell (e.g. \citealp{Jencson_2021}), so this is consistent with the low $t_{2}$ of TUVO-22albb. \

As discussed Sect. \ref{introduction}, very fast nova light curves are indicative of massive WDs. While this general connection is well-established (e.g. \citealp{Hachisu_2016,Shara_2017,Hachisu_2018,Darnley_2021}), the variety of different parameters of a nova that affect the light curve\footnote{\tiny For example, the chemical composition of the WD and of the accreted matter, internal shocks in the ejecta, dust formation in the ejecta, and interactions of the ejecta with the giant's wind in the case of a symbiotic. See \citep{Chomiuk_2021} and references therein for a discussion.} imply that obtaining precise estimates of the WD mass based solely on the speed class is difficult. Nonetheless, useful first approximations can be inferred. Very fast novae should harbour massive WDs of at least $1.25$ M$_{\odot}$ \citep{Yaron_2005,Hillman_2016,Shara_2017}, which gives a likely lower limit for the WD of TUVO-22albb. By comparing $t_{2}$ for TUVO-22albb to those of other very fast novae in symbiotics (e.g. T CrB with $t_{2}$=4 d and M$_{WD}$=1.38 M$_{\odot}$; V1534 with $t_{2}$=5.6 d and M$_{WD}$=1.37 M$_{\odot}$; V745 Sco with $t_{2}$=2 d and M$_{WD}$=1.38 M$_{\odot}$; \citealp{Hachisu_2018}), we suspect that the mass of the WD is in fact at least 1.35 M$_{\odot}$, which is near the maximum possible mass for a WD. \

The progenitors of Type Ia SNe, which are still not known, must have WD masses close to the Chandrasekhar mass limit of $\sim$1.4 M$_{\odot}$ \citep{Nomoto_1982}. Very fast novae already have WDs with masses close to this ($>$1.25 M$_{\odot}$), but to be viable Type Ia SN progenitor candidates, their WDs must be able to grow further. \citet{Hillman_2016} show that WDs in novae can grow in mass to the Chandrasekhar limit for a large range of accretion rates and initial WD masses. The very fast novae which originate in symbiotics may represent an even more likely channel to Type Ia SNe than those in CVs \citep{Munari_1992,Patat_2011,Darnley_2011}. This may be because the presence of the evolved giant in symbiotics results in a higher accretion rate \citep{Darnley_2011}, which allows the WD to grow more quickly (assuming not all the accreted material is ejected; e.g. \citealp{Hillman_2016}). The very fast nature of TUVO-22albb and consequently its high inferred WD mass, together with its likely origin in a symbiotic, overall make it an optimal candidate for the Type Ia SN progenitor channel. \

\begin{acknowledgements}
DM is partly supported by the Netherlands Research School for Astronomy (NOVA). \\

We acknowledge the use of public data from the Swift data archive. \\

This research has made use of the VizieR catalogue access tool, CDS, Strasbourg, France (DOI : 10.26093/cds/vizier), and NASA's Astrophysics Data System Bibliographic Services. \\

We thank the referee greatly for the very helpful and insightful suggestions to improve this paper.

\end{acknowledgements}

\bigskip

\bibliography{mybibliography.bib} 

\begin{thebibliography}{62}
\expandafter\ifx\csname natexlab\endcsname\relax\def\natexlab#1{#1}\fi

\bibitem[{{Allen}(1980)}]{Allen_1980}
{Allen}, D.~A. 1980, \mnras, 192, 521

\bibitem[{{Bode} \& {Evans}(2008)}]{Bode_2008}
{Bode}, M.~F. \& {Evans}, A. 2008, {Classical Novae}, Vol.~43

\bibitem[{{Bozzo} {et~al.}(2022){Bozzo}, {Romano}, {Ferrigno}, \&
  {Oskinova}}]{Bozzo_2022}
{Bozzo}, E., {Romano}, P., {Ferrigno}, C., \& {Oskinova}, L. 2022, \mnras, 513,
  42

\bibitem[{{Breeveld} {et~al.}(2010){Breeveld}, {Curran}, {Hoversten}, {Koch},
  {Landsman}, {Marshall}, {Page}, {Poole}, {Roming}, {Smith}, {Still},
  {Yershov}, {Blustin}, {Brown}, {Gronwall}, {Holland}, {Kuin}, {McGowan},
  {Rosen}, {Boyd}, {Broos}, {Carter}, {Chester}, {Hancock}, {Huckle}, {Immler},
  {Ivanushkina}, {Kennedy}, {Mason}, {Morgan}, {Oates}, {de Pasquale},
  {Schady}, {Siegel}, \& {vand en Berk}}]{Breeveld_2010}
{Breeveld}, A.~A., {Curran}, P.~A., {Hoversten}, E.~A., {et~al.} 2010, \mnras,
  406, 1687

\bibitem[{{Brown} {et~al.}(2010){Brown}, {Roming}, {Milne}, {Bufano},
  {Ciardullo}, {Elias-Rosa}, {Filippenko}, {Foley}, {Gehrels}, {Gronwall},
  {Hicken}, {Holland}, {Hoversten}, {Immler}, {Kirshner}, {Li}, {Mazzali},
  {Phillips}, {Pritchard}, {Still}, {Turatto}, \& {Vanden Berk}}]{Brown_2010}
{Brown}, P.~J., {Roming}, P. W.~A., {Milne}, P., {et~al.} 2010, \apj, 721, 1608

\bibitem[{{Burrows} {et~al.}(2005){Burrows}, {Hill}, {Nousek}, {Kennea},
  {Wells}, {Osborne}, {Abbey}, {Beardmore}, {Mukerjee}, {Short}, {Chincarini},
  {Campana}, {Citterio}, {Moretti}, {Pagani}, {Tagliaferri}, {Giommi},
  {Capalbi}, {Tamburelli}, {Angelini}, {Cusumano}, {Br{\"a}uninger}, {Burkert},
  \& {Hartner}}]{Burrows_2005}
{Burrows}, D.~N., {Hill}, J.~E., {Nousek}, J.~A., {et~al.} 2005, \ssr, 120, 165

\bibitem[{{Cao} {et~al.}(2012){Cao}, {Kasliwal}, {Neill}, {Kulkarni}, {Lou},
  {Ben-Ami}, {Bloom}, {Cenko}, {Law}, {Nugent}, {Ofek}, {Poznanski}, \&
  {Quimby}}]{Cao_2012}
{Cao}, Y., {Kasliwal}, M.~M., {Neill}, J.~D., {et~al.} 2012, \apj, 752, 133

\bibitem[{{Chomiuk} {et~al.}(2021){Chomiuk}, {Metzger}, \&
  {Shen}}]{Chomiuk_2021}
{Chomiuk}, L., {Metzger}, B.~D., \& {Shen}, K.~J. 2021, \araa, 59
  [\eprint[arXiv]{2011.08751}]

\bibitem[{{Cioni} {et~al.}(2006){Cioni}, {Girardi}, {Marigo}, \&
  {Habing}}]{Cioni_2006}
{Cioni}, M. R.~L., {Girardi}, L., {Marigo}, P., \& {Habing}, H.~J. 2006, \aap,
  448, 77

\bibitem[{{Darnley}(2021)}]{Darnley_2021}
{Darnley}, M.~J. 2021, in The Golden Age of Cataclysmic Variables and Related
  Objects V, Vol. 2-7, 44

\bibitem[{{Darnley} {et~al.}(2016){Darnley}, {Henze}, {Bode}, {Hachisu},
  {Hernanz}, {Hornoch}, {Hounsell}, {Kato}, {Ness}, {Osborne}, {Page},
  {Ribeiro}, {Rodr{\'\i}guez-Gil}, {Shafter}, {Shara}, {Steele}, {Williams},
  {Arai}, {Arcavi}, {Barsukova}, {Boumis}, {Chen}, {Fabrika}, {Figueira},
  {Gao}, {Gehrels}, {Godon}, {Goranskij}, {Harman}, {Hartmann}, {Hosseinzadeh},
  {Horst}, {Itagaki}, {Jos{\'e}}, {Kabashima}, {Kaur}, {Kawai}, {Kennea},
  {Kiyota}, {Ku{\v{c}}{\'a}kov{\'a}}, {Lau}, {Maehara}, {Naito}, {Nakajima},
  {Nishiyama}, {O'Brien}, {Quimby}, {Sala}, {Sano}, {Sion}, {Valeev},
  {Watanabe}, {Watanabe}, {Williams}, \& {Xu}}]{Darnley_2016}
{Darnley}, M.~J., {Henze}, M., {Bode}, M.~F., {et~al.} 2016, \apj, 833, 149

\bibitem[{{Darnley} {et~al.}(2011){Darnley}, {Ribeiro}, {Bode}, \&
  {Munari}}]{Darnley_2011}
{Darnley}, M.~J., {Ribeiro}, V.~A.~R.~M., {Bode}, M.~F., \& {Munari}, U. 2011,
  \aap, 530, A70

\bibitem[{{Evans} {et~al.}(2009){Evans}, {Beardmore}, {Page}, {Osborne},
  {O'Brien}, {Willingale}, {Starling}, {Burrows}, {Godet}, {Vetere}, {Racusin},
  {Goad}, {Wiersema}, {Angelini}, {Capalbi}, {Chincarini}, {Gehrels}, {Kennea},
  {Margutti}, {Morris}, {Mountford}, {Pagani}, {Perri}, {Romano}, \&
  {Tanvir}}]{Evans_2009}
{Evans}, P.~A., {Beardmore}, A.~P., {Page}, K.~L., {et~al.} 2009, \mnras, 397,
  1177

\bibitem[{{Fazio} {et~al.}(2004){Fazio}, {Hora}, {Allen}, {Ashby}, {Barmby},
  {Deutsch}, {Huang}, {Kleiner}, {Marengo}, {Megeath}, {Melnick}, {Pahre},
  {Patten}, {Polizotti}, {Smith}, {Taylor}, {Wang}, {Willner}, {Hoffmann},
  {Pipher}, {Forrest}, {McMurty}, {McCreight}, {McKelvey}, {McMurray}, {Koch},
  {Moseley}, {Arendt}, {Mentzell}, {Marx}, {Losch}, {Mayman}, {Eichhorn},
  {Krebs}, {Jhabvala}, {Gezari}, {Fixsen}, {Flores}, {Shakoorzadeh}, {Jungo},
  {Hakun}, {Workman}, {Karpati}, {Kichak}, {Whitley}, {Mann}, {Tollestrup},
  {Eisenhardt}, {Stern}, {Gorjian}, {Bhattacharya}, {Carey}, {Nelson},
  {Glaccum}, {Lacy}, {Lowrance}, {Laine}, {Reach}, {Stauffer}, {Surace},
  {Wilson}, {Wright}, {Hoffman}, {Domingo}, \& {Cohen}}]{Fazio_2004}
{Fazio}, G.~G., {Hora}, J.~L., {Allen}, L.~E., {et~al.} 2004, \apjs, 154, 10

\bibitem[{{Gallagher} \& {Starrfield}(1978)}]{Gallagher_1978}
{Gallagher}, J.~S. \& {Starrfield}, S. 1978, \araa, 16, 171

\bibitem[{{Gehrels}(1986)}]{Gehrels_1986}
{Gehrels}, N. 1986, \apj, 303, 336

\bibitem[{Gehrels {et~al.}(2004)Gehrels, Chincarini, Giommi, Mason, Nousek,
  Wells, White, Barthelmy, Burrows, Cominsky, Hurley, Marshall, Meszaros,
  Roming, Angelini, Barbier, Belloni, Campana, Caraveo, Chester, Citterio,
  Cline, Cropper, Cummings, Dean, Feigelson, Fenimore, Frail, Fruchter,
  Garmire, Gendreau, Ghisellini, Greiner, Hill, Hunsberger, Krimm, Kulkarni,
  Kumar, Lebrun, Lloyd-Ronning, Markwardt, Mattson, Mushotzky, Norris, Osborne,
  Paczynski, Palmer, Park, Parsons, Paul, Rees, Reynolds, Rhoads, Sasseen,
  Schaefer, Short, Smale, Smith, Stella, Tagliaferri, Takahashi, Tashiro,
  Townsley, Tueller, Turner, Vietri, Voges, Ward, Willingale, Zerbi, \&
  Zhang}]{Gehrels_2004}
Gehrels, N., Chincarini, G., Giommi, P., {et~al.} 2004, \apj, 611, 1005

\bibitem[{{Gieren} {et~al.}(2005){Gieren}, {Pietrzy{\'n}ski}, {Soszy{\'n}ski},
  {Bresolin}, {Kudritzki}, {Minniti}, \& {Storm}}]{Gieren_2005}
{Gieren}, W., {Pietrzy{\'n}ski}, G., {Soszy{\'n}ski}, I., {et~al.} 2005, \apj,
  628, 695

\bibitem[{{G{\"u}nther} {et~al.}(2020){G{\"u}nther}, {Zhan}, {Seager},
  {Rimmer}, {Ranjan}, {Stassun}, {Oelkers}, {Daylan}, {Newton}, {Kristiansen},
  {Olah}, {Gillen}, {Rappaport}, {Ricker}, {Vanderspek}, {Latham}, {Winn},
  {Jenkins}, {Glidden}, {Fausnaugh}, {Levine}, {Dittmann}, {Quinn},
  {Krishnamurthy}, \& {Ting}}]{Gunther_2020}
{G{\"u}nther}, M.~N., {Zhan}, Z., {Seager}, S., {et~al.} 2020, \aj, 159, 60

\bibitem[{{Hachisu} \& {Kato}(2000)}]{Hachisu_2000}
{Hachisu}, I. \& {Kato}, M. 2000, \apj, 540, 447

\bibitem[{{Hachisu} \& {Kato}(2016)}]{Hachisu_2016}
{Hachisu}, I. \& {Kato}, M. 2016, \apjs, 223, 21

\bibitem[{{Hachisu} \& {Kato}(2018)}]{Hachisu_2018}
{Hachisu}, I. \& {Kato}, M. 2018, \apjs, 237, 4

\bibitem[{{Hachisu} {et~al.}(2000){Hachisu}, {Kato}, {Kato}, \&
  {Matsumoto}}]{Hachisu_2000_2}
{Hachisu}, I., {Kato}, M., {Kato}, T., \& {Matsumoto}, K. 2000, \apjl, 528, L97

\bibitem[{{Hillman} {et~al.}(2016){Hillman}, {Prialnik}, {Kovetz}, \&
  {Shara}}]{Hillman_2016}
{Hillman}, Y., {Prialnik}, D., {Kovetz}, A., \& {Shara}, M.~M. 2016, \apj, 819,
  168

\bibitem[{{I{\l}kiewicz} {et~al.}(2016){I{\l}kiewicz}, {Miko{\l}ajewska},
  {Stoyanov}, {Manousakis}, \& {Miszalski}}]{Ilkiewicz_2016}
{I{\l}kiewicz}, K., {Miko{\l}ajewska}, J., {Stoyanov}, K., {Manousakis}, A., \&
  {Miszalski}, B. 2016, \mnras, 462, 2695

\bibitem[{{Jencson} {et~al.}(2021){Jencson}, {Andrews}, {Bond}, {Karambelkar},
  {Sand}, {van Dyk}, {Blagorodnova}, {Boyer}, {Kasliwal}, {Lau}, {Mohamed},
  {Williams}, {Whitelock}, {Amaro}, {Bostroem}, {Dong}, {Lundquist}, {Valenti},
  {Wyatt}, {Burke}, {De}, {Jha}, {Johansson}, {Rojas-Bravo}, {Coulter},
  {Foley}, {Gehrz}, {Haislip}, {Hiramatsu}, {Howell}, {Kilpatrick}, {Masci},
  {McCully}, {Ngeow}, {Pan}, {Pellegrino}, {Piro}, {Kouprianov}, {Reichart},
  {Rest}, {Rest}, \& {Smith}}]{Jencson_2021}
{Jencson}, J.~E., {Andrews}, J.~E., {Bond}, H.~E., {et~al.} 2021, \apj, 920,
  127

\bibitem[{{Joshi} {et~al.}(2015){Joshi}, {Banerjee}, {Ashok}, {Venkataraman},
  \& {Walter}}]{Joshi_2015}
{Joshi}, V., {Banerjee}, D.~P.~K., {Ashok}, N.~M., {Venkataraman}, V., \&
  {Walter}, F.~M. 2015, \mnras, 452, 3696

\bibitem[{{Kuijken}(2011)}]{Kuijken_2011}
{Kuijken}, K. 2011, The Messenger, 146, 8

\bibitem[{{Lasota}(2001)}]{Lasota_2001}
{Lasota}, J.-P. 2001, \nar, 45, 449

\bibitem[{{Maoz} {et~al.}(2014){Maoz}, {Mannucci}, \& {Nelemans}}]{Maoz_2014}
{Maoz}, D., {Mannucci}, F., \& {Nelemans}, G. 2014, \araa, 52, 107

\bibitem[{{Mason} {et~al.}(2012){Mason}, {Ederoclite}, {Williams}, {Della
  Valle}, \& {Setiawan}}]{Mason_2012}
{Mason}, E., {Ederoclite}, A., {Williams}, R.~E., {Della Valle}, M., \&
  {Setiawan}, J. 2012, \aap, 544, A149

\bibitem[{{Mikolajewska}(2010)}]{Mikolajewska_2010}
{Mikolajewska}, J. 2010, arXiv e-prints, arXiv:1011.5657

\bibitem[{{Miko{\l}ajewska}(2012)}]{Mikolajewska_2012}
{Miko{\l}ajewska}, J. 2012, Baltic Astronomy, 21, 5

\bibitem[{{Modiano} {et~al.}(2020){Modiano}, {Parikh}, \&
  {Wijnands}}]{Modiano_2020}
{Modiano}, D., {Parikh}, A.~S., \& {Wijnands}, R. 2020, \aap, 634, A132

\bibitem[{{Modiano} {et~al.}(2022){Modiano}, {Wijnands}, {Parikh}, {van
  Opijnen}, {Verberne}, \& {van Etten}}]{Modiano_2022}
{Modiano}, D., {Wijnands}, R., {Parikh}, A., {et~al.} 2022, \aap, 663, A5

\bibitem[{{Munari}(2019)}]{Munari_2019}
{Munari}, U. 2019, arXiv e-prints, arXiv:1909.01389

\bibitem[{{Munari} {et~al.}(2011){Munari}, {Joshi}, {Ashok}, {Banerjee},
  {Valisa}, {Milani}, {Siviero}, {Dallaporta}, \& {Castellani}}]{Munari_2011}
{Munari}, U., {Joshi}, V.~H., {Ashok}, N.~M., {et~al.} 2011, \mnras, 410, L52

\bibitem[{{Munari} \& {Renzini}(1992)}]{Munari_1992}
{Munari}, U. \& {Renzini}, A. 1992, \apjl, 397, L87

\bibitem[{{Munari} \& {Valisa}(2022)}]{Munari_2022}
{Munari}, U. \& {Valisa}, P. 2022, arXiv e-prints, arXiv:2203.01378

\bibitem[{{Nomoto}(1982)}]{Nomoto_1982}
{Nomoto}, K. 1982, \apj, 253, 798

\bibitem[{{Page} {et~al.}(2022){Page}, {Beardmore}, {Osborne}, {Munari},
  {Ness}, {Evans}, {Bode}, {Darnley}, {Drake}, {Kuin}, {O'Brien}, {Orio},
  {Shore}, {Starrfield}, \& {Woodward}}]{Page_2022}
{Page}, K.~L., {Beardmore}, A.~P., {Osborne}, J.~P., {et~al.} 2022, \mnras,
  514, 1557

\bibitem[{{Page} {et~al.}(2020){Page}, {Kuin}, {Beardmore}, {Walter},
  {Osborne}, {Markwardt}, {Ness}, {Orio}, \& {Sokolovsky}}]{Page_2020}
{Page}, K.~L., {Kuin}, N.~P.~M., {Beardmore}, A.~P., {et~al.} 2020, \mnras,
  499, 4814

\bibitem[{{Page} {et~al.}(2015){Page}, {Osborne}, {Kuin}, {Henze}, {Walter},
  {Beardmore}, {Bode}, {Darnley}, {Delgado}, {Drake}, {Hernanz}, {Mukai},
  {Nelson}, {Ness}, {Schwarz}, {Shore}, {Starrfield}, \&
  {Woodward}}]{Page_2015}
{Page}, K.~L., {Osborne}, J.~P., {Kuin}, N.~P.~M., {et~al.} 2015, \mnras, 454,
  3108

\bibitem[{{Patat} {et~al.}(2011){Patat}, {Chugai}, {Podsiadlowski}, {Mason},
  {Melo}, \& {Pasquini}}]{Patat_2011}
{Patat}, F., {Chugai}, N.~N., {Podsiadlowski}, P., {et~al.} 2011, \aap, 530,
  A63

\bibitem[{Payne-Gaposchkin(1964)}]{payne1964galactic}
Payne-Gaposchkin, C. 1964, The Galactic Novae, DOVER BOOKS ON ASTRONOMY AND
  ASTROPHYSICS (Dover Publications)

\bibitem[{{Poole} {et~al.}(2008){Poole}, {Breeveld}, {Page}, {Landsman},
  {Holland}, {Roming}, {Kuin}, {Brown}, {Gronwall}, {Hunsberger}, {Koch},
  {Mason}, {Schady}, {vanden Berk}, {Blustin}, {Boyd}, {Broos}, {Carter},
  {Chester}, {Cucchiara}, {Hancock}, {Huckle}, {Immler}, {Ivanushkina},
  {Kennedy}, {Marshall}, {Morgan}, {Pandey}, {de Pasquale}, {Smith}, \&
  {Still}}]{Poole_2008}
{Poole}, T.~S., {Breeveld}, A.~A., {Page}, M.~J., {et~al.} 2008, \mnras, 383,
  627

\bibitem[{{Roming} {et~al.}(2005){Roming}, {Kennedy}, {Mason}, {Nousek}, {Ahr},
  {Bingham}, {Broos}, {Carter}, {Hancock}, {Huckle}, {Hunsberger}, {Kawakami},
  {Killough}, {Koch}, {McLelland}, {Smith}, {Smith}, {Soto}, {Boyd},
  {Breeveld}, {Holland}, {Ivanushkina}, {Pryzby}, {Still}, \&
  {Stock}}]{Roming_2005}
{Roming}, P. W.~A., {Kennedy}, T.~E., {Mason}, K.~O., {et~al.} 2005, \ssr, 120,
  95

\bibitem[{{Schaefer}(2010)}]{Schaefer_2010}
{Schaefer}, B.~E. 2010, \apjs, 187, 275

\bibitem[{{Sekiguchi} {et~al.}(1990){Sekiguchi}, {Whitelock}, {Feast},
  {Barrett}, {Caldwell}, {Carter}, {Catchpole}, {Laing}, {Laney}, {Marang}, \&
  {van Wyck}}]{Sekiguchi_1990}
{Sekiguchi}, K., {Whitelock}, P.~A., {Feast}, M.~W., {et~al.} 1990, \mnras,
  246, 78

\bibitem[{{Shanks} {et~al.}(2015){Shanks}, {Metcalfe}, {Chehade}, {Findlay},
  {Irwin}, {Gonzalez-Solares}, {Lewis}, {Yoldas}, {Mann}, {Read}, {Sutorius},
  \& {Voutsinas}}]{Shanks_2015}
{Shanks}, T., {Metcalfe}, N., {Chehade}, B., {et~al.} 2015, \mnras, 451, 4238

\bibitem[{{Shara} {et~al.}(2018){Shara}, {Prialnik}, {Hillman}, \&
  {Kovetz}}]{Shara_2017}
{Shara}, M.~M., {Prialnik}, D., {Hillman}, Y., \& {Kovetz}, A. 2018, \apj, 860,
  110

\bibitem[{{Shibayama} {et~al.}(2013){Shibayama}, {Maehara}, {Notsu}, {Notsu},
  {Nagao}, {Honda}, {Ishii}, {Nogami}, \& {Shibata}}]{Shibayama_2013}
{Shibayama}, T., {Maehara}, H., {Notsu}, S., {et~al.} 2013, \apjs, 209, 5

\bibitem[{{Siegel} {et~al.}(2012){Siegel}, {Hoversten}, {Bond}, {Stark}, \&
  {Breeveld}}]{Siegel_2012}
{Siegel}, M.~H., {Hoversten}, E., {Bond}, H.~E., {Stark}, M., \& {Breeveld},
  A.~A. 2012, \aj, 144, 65

\bibitem[{{Siegel} {et~al.}(2014){Siegel}, {Porterfield}, {Linevsky}, {Bond},
  {Holland}, {Hoversten}, {Berrier}, {Breeveld}, {Brown}, \&
  {Gronwall}}]{Siegel_2014}
{Siegel}, M.~H., {Porterfield}, B.~L., {Linevsky}, J.~S., {et~al.} 2014, \aj,
  148, 131

\bibitem[{{Starrfield} {et~al.}(2016){Starrfield}, {Iliadis}, \&
  {Hix}}]{Starrfield_2016}
{Starrfield}, S., {Iliadis}, C., \& {Hix}, W.~R. 2016, \pasp, 128, 051001

\bibitem[{{Starrfield} {et~al.}(1972){Starrfield}, {Truran}, {Sparks}, \&
  {Kutter}}]{Starrfield_1972}
{Starrfield}, S., {Truran}, J.~W., {Sparks}, W.~M., \& {Kutter}, G.~S. 1972,
  \apj, 176, 169

\bibitem[{{Strope} {et~al.}(2010){Strope}, {Schaefer}, \&
  {Henden}}]{Strope_2010}
{Strope}, R.~J., {Schaefer}, B.~E., \& {Henden}, A.~A. 2010, \aj, 140, 34

\bibitem[{{Sutherland} {et~al.}(2015){Sutherland}, {Emerson}, {Dalton},
  {Atad-Ettedgui}, {Beard}, {Bennett}, {Bezawada}, {Born}, {Caldwell}, {Clark},
  {Craig}, {Henry}, {Jeffers}, {Little}, {McPherson}, {Murray}, {Stewart},
  {Stobie}, {Terrett}, {Ward}, {Whalley}, \& {Woodhouse}}]{Sutherland_2015}
{Sutherland}, W., {Emerson}, J., {Dalton}, G., {et~al.} 2015, \aap, 575, A25

\bibitem[{{Warner}(1995)}]{Warner_1995}
{Warner}, B. 1995, {Cataclysmic variable stars}, Vol.~28

\bibitem[{{Werner} {et~al.}(2004){Werner}, {Roellig}, {Low}, {Rieke}, {Rieke},
  {Hoffmann}, {Young}, {Houck}, {Brandl}, {Fazio}, {Hora}, {Gehrz}, {Helou},
  {Soifer}, {Stauffer}, {Keene}, {Eisenhardt}, {Gallagher}, {Gautier}, {Irace},
  {Lawrence}, {Simmons}, {Van Cleve}, {Jura}, {Wright}, \&
  {Cruikshank}}]{Werner_2004}
{Werner}, M.~W., {Roellig}, T.~L., {Low}, F.~J., {et~al.} 2004, \apjs, 154, 1

\bibitem[{{Whitelock} {et~al.}(2009){Whitelock}, {Menzies}, {Feast},
  {Matsunaga}, {Tanab{\'e}}, \& {Ita}}]{Whitelock_2009}
{Whitelock}, P.~A., {Menzies}, J.~W., {Feast}, M.~W., {et~al.} 2009, \mnras,
  394, 795

\bibitem[{{Yaron} {et~al.}(2005){Yaron}, {Prialnik}, {Shara}, \&
  {Kovetz}}]{Yaron_2005}
{Yaron}, O., {Prialnik}, D., {Shara}, M.~M., \& {Kovetz}, A. 2005, \apj, 623,
  398

\end{thebibliography}

\newpage
\begin{appendix}

\section{Long-term UVOT light curve} \label{appendix_pv}

\begin{figure*}[b]
    \centering
    \includegraphics[width=0.8\textwidth]{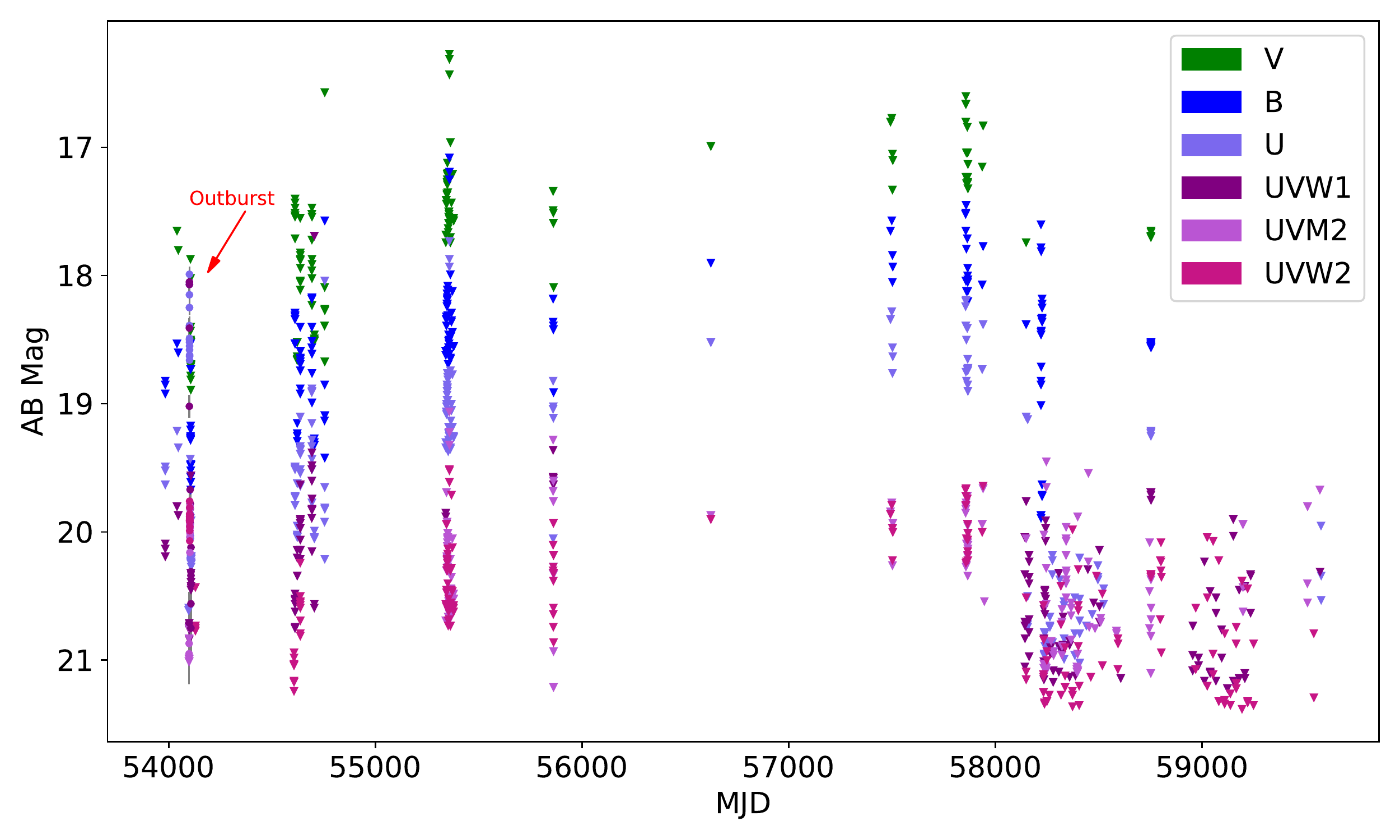}
    \caption{The long-term UVOT light curve of TUVO-22albb in six filters as automatically generated by our pipeline (for details about the pipeline see \citealp{Modiano_2022}). Triangles indicate upper limits. The outburst which is the focus of this study is labelled; no additional outbursts were detected. }
    \label{fig:pv_appendix}
\end{figure*}

In Fig. \ref{fig:pv_appendix} we provide the long-term UVOT light curve of TUVO-22albb. Details of how this light curve is created are described in (\citealp{Modiano_2022}; Sect. 2.6), and we refer to the reproduction package of this paper for further details. The light curve only shows one outburst detected from the source. This outburst is extensively discussed in the main text of the paper.

\section{Co-addition of exposures} \label{appendix_coadd}

Here we provide details for the observations used to create the light curve shown in Figs. \ref{fig:lightcurve} and \ref{fig:lc_exp_fits}. Table \ref{tab:obs_used} summarises the  observations that were used to study the outburst of the source.\ 

As discussed in Sect. \ref{outburst_photometry}, some exposures in which there was no clear detection of the source were co-added. We chose the intervals over which to co-add exposures such that each co-added exposure resulted in a clear detection. The co-addition of exposures and the intervals that were co-added are detailed in Table \ref{tab:obs_used}. Co-addition was always performed independently for each filter, using the HEASoft (v6.29) tool \texttt{uvotimsum}\footnote{\tiny\url{https://heasarc.gsfc.nasa.gov/lheasoft/ftools/headas/uvotimsum.html}}.\

For the two pre-outburst observations (ObsIDs 00030848002 and 00030848003), we co-added all exposures to check for a detection. Since no detection was made in either filter in the resulting co-added exposures, in Figs. \ref{fig:lightcurve} and \ref{fig:lc_exp_fits} and in Table \ref{tab:obs_used} we only include the individual exposures for those observations. \

For all co-added exposures, the time stamps used for the light curve shown in Fig. \ref{fig:lightcurve} are the average of the time stamps of their constituent exposures, weighted by exposure time. The standard deviation on the time stamps (also weighted by exposure time) was used as the error on the time. For the exposures that were not co-added, we define the error bars on the time as the exposure times. However, these were not included in Fig. \ref{fig:lightcurve} because they were much smaller than the size of the markers. \

\section{Photometry} \label{appendix_photometry}

Here we provide details about how we optimised the \texttt{uvotsource} photometry in order to obtain the most accurate results. We refer to the reproduction package of this paper for all the files and details needed to perform the photometry as described. We additionally provide a table of all the magnitudes measured from the UVOT data for all images in which a detection was made (Table \ref{tab:appendix_mags_final}). The magnitudes given were converted from fluxes after the corrections described in Sect. \ref{outburst_photometry}. \

The aperture used for photometry was selected as a circular region of radius \SI{3.5}{\arcsecond} centred at the position of the transient. The aperture was checked by eye for each exposure, and the centre was manually adjusted if needed to account for small-scale misalignments between exposures (of $\sim$0.5$\arcsec$; \citealp{Poole_2008}). The radius of the aperture was chosen so as to optimise the signal-to-noise (SNR) on the data during the outburst peak (and then not modified for all other observations for consistency). Specifically, this was done in order to keep contamination from sources close to the transient position minimal. We note however that a smaller aperture radius of 1.5$\arcsec$ was tested and had almost no effect on the photometry, because \texttt{uvotsource} corrects for aperture size. \ 
However, the measured fluxes still suffer from contamination from the high background emission at the source position (due to unresolved stars and diffuse emission objects in NGC 300; see Fig. \ref{fig:uvot_images}). To take this effect into account, we used the same method described in \citet{Modiano_2020}, who encountered a similar issue with diffuse emission in the globular cluster 47 Tuc. The method consists of determining the flux at the source position by carrying out photometry using the same source aperture, but during quiescence (i.e. when the source was not detected; see Sect. \ref{results_quiescence}). This `contamination flux' is equal to the flux from other sources in NGC 300 at the source position. For each measured flux of the transient during the outburst, we subtracted this contamination flux, independently for each filter. \ 

The instrumental background is accounted for automatically by \texttt{uvotsource} using a user-provided background region. We selected a background region near the edge of the exposures (far from the centre of NGC 300), to avoid over-correcting (because the galaxy background is already accounted for in our method). \ 

For the pre-outburst $UVM2$ and $UVW1$ exposures, no source was detected in any of the exposures nor in the co-added exposures. We therefore reported the photometry on these data (in Fig. \ref{fig:lightcurve}) for individual exposures. The photometry for these exposures was performed in the same exact way as for the detections, including the removal of the diffuse emission. Therefore they are all exhibit roughly zero flux (minor departures from zero, including to negative fluxes, are likely due to small-scale variations in the diffuse emission at the source position). To represent these as non-detections we, display them as open circles in Fig. \ref{fig:lightcurve}.\ 

To determine upper limits from the co-added quiescence exposures built from the entire archival UVOT data set of NGC 300 (except the outburst observations), we used a smaller aperture size of \SI{1.2}{\arcsecond}, to avoid contamination from bright objects close to the source position.

\section{XRT data} \label{appendix_xrt}

Here we provide details regarding the XRT data analysed. We used the online XRT analysis tool\footnote{\tiny\url{https://www.swift.ac.uk/user_objects/}} developed by \citet{Evans_2009} to build combined X-ray images of the field of TUVO-22albb in the energy ranges 0.3--10 keV, 0.3--1.5 keV, and 1.5--10 keV. We combined the outburst observations (ObsIDs 00030848001--00030848012; all in photon counting, PC, mode), resulting in a total of $\sim$84 ks of exposure time. We separately combined all quiescence observations (all other ObsIDs; also all in PC mode), which accounted for $\sim$500 ks of exposure time. No detections were made at the source position in either the combined outburst images or the combined quiescent images (in any of the energy ranges). We estimated upper limits using the Webpimms\footnote{\tiny\url{https://heasarc.gsfc.nasa.gov/cgi-bin/Tools/w3pimms/w3pimms.pl}} tool available from HEASARC, by inputting the predicted column density N${_H}$ (at the source position) obtained using the N${_H}$ tool\footnote{\tiny\url{https://heasarc.gsfc.nasa.gov/cgi-bin/Tools/w3nh/w3nh.pl}}, an expected power law index of 1.8 (similar to other quiescent symbiotics; see e.g. \citealp{Bozzo_2022}), and the 3-$\sigma$ confidence interval prescriptions from \citet{Gehrels_1986}. We report unabsorbed X-ray 0.3--10 keV flux upper limits of 3.7$\times$10$^{-15}$ erg cm$^{-2}$ s$^{-1}$ and 6.2$\times$10$^{-16}$ erg cm$^{-2}$ s$^{-1}$ for outburst and quiescence, respectively. Using the distance to NGC 300 of 1.88 Mpc, we report upper limits on the outburst and quiescence X-ray luminosity of 1.6$\times$10$^{36}$ erg s$^{-1}$ and 2.6$\times$10$^{35}$ erg s$^{-1}$, respectively. \

\begin{figure}
    \centering
    \includegraphics[width=0.3\textwidth]{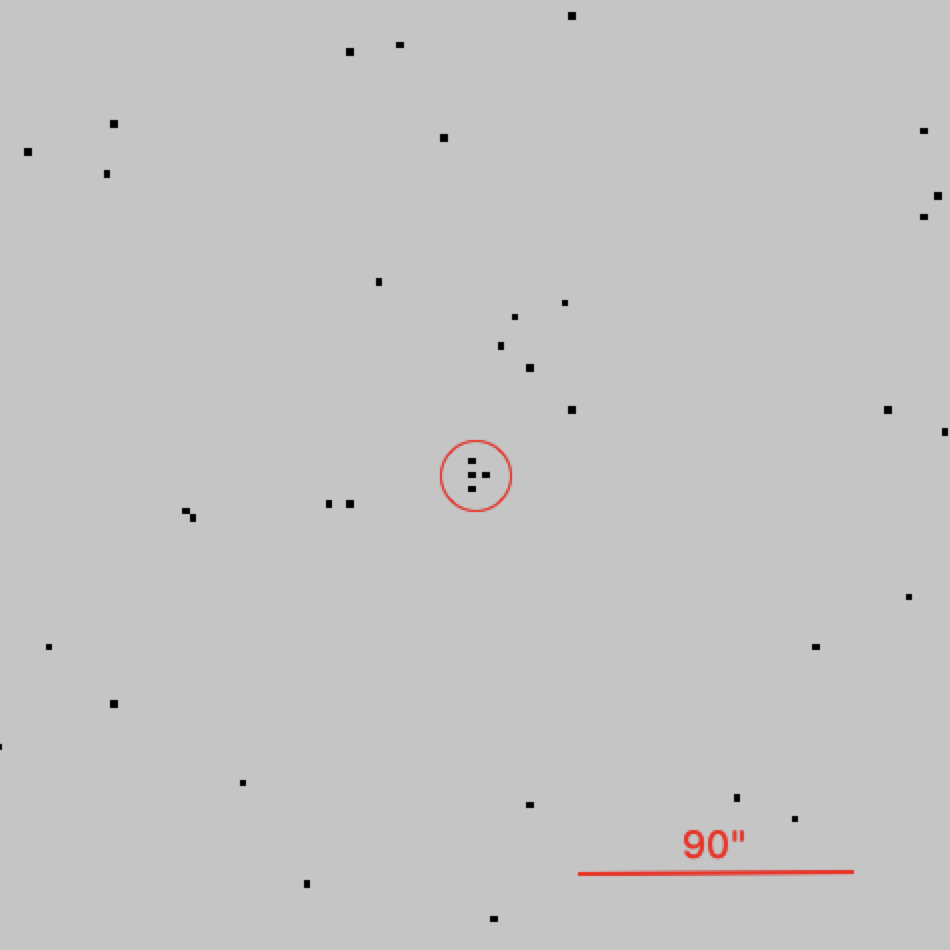}
    \caption{Part of the XRT image (ObsID 00035866004) in which a possible faint detection was made at the position of TUVO-22albb (labelled with a red circle). North is up and east is left.}
    \label{fig:xrt_image}
\end{figure}

We inspected all individual XRT images during outburst by eye, and found no detections. We also checked the observation carried out $\sim$30 days after the outburst peak (ObsID 00035866004), and found a possible, faint detection (SNR < 2), shown in Fig. \ref{fig:xrt_image}. For the fluxes derived from this tentative detection, we used the same inputs as above (power-law index and N${_H}$), together with the measured count rate, to determine the flux (see Sect. \ref{xrt_analysis}). We did not check any subsequent XRT data by eye (all further observations were obtained over a year later).

\begin{table*}[b]
    \centering
    \caption{UVOT observations used to construct the light curve shown in Fig. \ref{fig:lightcurve}.}
    \small
    \begin{tabular}{cccccc}
    \toprule
        Filter & ObsID & No. of exposures & Total exp. time (s) & Co-added exposures? & Notes\\
        \hline
        $UVM2$ & 00030848002 & 10 & 9982 & No & Source never detected$^{*}$ \\
        $UVW1$ & 00030848003 & 15 & 9695 & No & Source never detected$^{*}$ \\
        $U$ & 00030848004 & 15 & 9511 & No & \\
        $UVW2$ & 00030848005 & 15 & 9333 & No & \\
        $UVM2$ & 00030848006 & 11 & 7798 & No & \\
        $UVW1$ & 00030848007 & 7 & 8969 & No & \\
        $U$ & 00030848008 & 7 & 2333 & No & \\
        $B$ & 00030848009,00030848010 & 11,11 & 2410,4184 & Yes & \\
        $V$ & 00030848009,00030848010 & 11,11 & 2725,4489 & Yes & \\
        $UVW1$ & 00030848011 & 11 & 7101 & Yes & \\
        $U$ & 00030848012 & 11 & 2061 & Yes & Co-added exposures 1--6 \\
        $U$ & 00030848012 & 11 & 1798 & Yes & Co-added exposures 7--11 \\
    \bottomrule
    \end{tabular}
    \parbox{0.9\textwidth}{\small%
    \vspace{1eX}
    $^{*}$ Co-addition was performed for this ObsID, but the source was not detected in any of the individual exposures nor in the co-added exposure. Therefore, in this table (and in Fig. \ref{fig:lightcurve}) the exposures are not co-added.}
    \label{tab:obs_used}
\end{table*}

\clearpage

\begin{table}[h]
    \centering
    \tiny
    \caption{Magnitudes calculated from UVOT fluxes.}
    \begin{tabular}{ccccc}
    \toprule
        MJD & Days after peak & Filter & Magnitude \\ \hline \hline
        
        54099.79 & -0.13 & $UVW1$ & 19.1 \\
        54099.86 & -0.07 & $UVW1$ & 18.4 \\
        54099.93 & 0.00 & $UVW1$ & 18.1 \\
        54099.99 & 0.07 & $UVW1$ & 18.1 \\
        
        54100.06 & 0.13 & $U$ & 18.0 \\
        54100.13 & 0.20 & $U$ & 18.0 \\
        54100.19 & 0.27 & $U$ & 18.1 \\
        54100.26 & 0.33 & $U$ & 18.2 \\
        54100.33 & 0.40 & $U$ & 18.2 \\
        54100.39 & 0.47 & $U$ & 18.3 \\
        54100.46 & 0.54 & $U$ & 18.4 \\
        54100.53 & 0.60 & $U$ & 18.5 \\
        54100.59 & 0.67 & $U$ & 18.5 \\
        54100.66 & 0.74 & $U$ & 18.4 \\
        54100.73 & 0.80 & $U$ & 18.5 \\
        54100.80 & 0.87 & $U$ & 18.4 \\
        54100.86 & 0.94 & $U$ & 18.6 \\
        54100.93 & 1.00 & $U$ & 18.6 \\
        54100.99 & 1.07 & $U$ & 18.5 \\
        
        54101.06 & 1.14 & $UVW2$ & 19.9 \\
        54101.13 & 1.21 & $UVW2$ & 20.1 \\
        54101.20 & 1.27 & $UVW2$ & 20.1 \\
        54101.26 & 1.34 & $UVW2$ & 19.9 \\
        54101.33 & 1.41 & $UVW2$ & 20.0 \\
        54101.40 & 1.47 & $UVW2$ & 20.1 \\
        54101.46 & 1.54 & $UVW2$ & 20.0 \\
        54101.53 & 1.61 & $UVW2$ & 20.0 \\
        54101.60 & 1.67 & $UVW2$ & 20.0 \\
        54101.67 & 1.74 & $UVW2$ & 20.0 \\
        54101.73 & 1.81 & $UVW2$ & 19.9 \\
        54101.80 & 1.87 & $UVW2$ & 20.1 \\
        54101.87 & 1.94 & $UVW2$ & 20.0 \\
        54101.93 & 2.01 & $UVW2$ & 20.0 \\
        54102.00 & 2.07 & $UVW2$ & 20.1 \\
        
        54102.07 & 2.14 & $UVM2$ & 20.1 \\
        54102.13 & 2.21 & $UVM2$ & 20.0 \\
        54102.20 & 2.28 & $UVM2$ & 20.1 \\
        54102.27 & 2.34 & $UVM2$ & 20.0 \\
        54102.34 & 2.41 & $UVM2$ & 20.1 \\
        54102.40 & 2.48 & $UVM2$ & 20.2 \\
        54102.47 & 2.54 & $UVM2$ & 20.2 \\
        54102.74 & 2.81 & $UVM2$ & 20.2 \\
        54102.80 & 2.88 & $UVM2$ & 20.1 \\
        54102.87 & 2.95 & $UVM2$ & 20.1 \\
        54102.93 & 3.01 & $UVM2$ & 20.0 \\
        
        54103.48 & 3.55 & $UVW1$ & 20.2 \\
        54103.54 & 3.62 & $UVW1$ & 20.0 \\
        54103.61 & 3.69 & $UVW1$ & 20.2 \\
        54103.68 & 3.75 & $UVW1$ & 20.2 \\
        54103.74 & 3.82 & $UVW1$ & 20.1 \\
        54103.81 & 3.89 & $UVW1$ & 20.1 \\
        54103.88 & 3.96 & $UVW1$ & 20.0 \\
        
        54104.22 & 4.29 & $U$ & 19.6 \\
        54104.29 & 4.36 & $U$ & 19.8 \\
        54104.35 & 4.43 & $U$ & 20.0 \\
        54104.42 & 4.50 & $U$ & 19.9 \\
        54104.49 & 4.56 & $U$ & 19.9 \\
        54104.55 & 4.63 & $U$ & 20.0 \\
        54104.94 & 5.02 & $U$ & 20.0 \\
        
        54106.15 & 6.22 & $V$ & 19.5 \\
        
        54106.16 & 6.24 & $B$ & 20.5 \\
        
        54107.42 & 7.49 & $UVW1$ & 20.9 \\
        
        54108.28 & 8.35 & $U$ & 20.5 \\
        54108.63 & 8.71 & $U$ & 20.6 \\
        
    \bottomrule
    \end{tabular}
    \parbox{0.45\textwidth}{\small%
    \vspace{1eX}
    \textbf{Notes. }For details about how the fluxes were determined, see Sect. \ref{outburst_photometry}. The peak time is defined as MJD=54099.90. All magnitudes are in the AB system. Errors on magnitudes are all on the order of 0.1--0.2 mags. For the co-added exposures, the MJD given is the average of the time stamps of each individual exposures, weighted by exposure time.}
    \label{tab:appendix_mags_final}
\end{table}

\newpage

\end{appendix}

\end{document}